\documentclass[aps]{revtex4}%
\usepackage{amsfonts}
\usepackage{amsmath}
\usepackage{amssymb}
\usepackage{graphicx}%

\begin{document}
\title{Crossing the phantom divide in brane cosmology with curvature corrections and
brane-bulk energy transfer}
\author{Shao-Feng Wu$^{1}$\footnote{Email: sfwu@shu.edu.cn}, Auttakit
Chatrabhuti$^{2}$\footnote{Email: auttakit@sc.chula.ac.th},
Guo-Hong Yang$^{1}$\footnote{Email: ghyang@mail.shu.edu.cn}, and
Peng-Ming Zhang$^{3,4}$\footnote{Email: zhpm@impcas.ac.cn}}
\affiliation{$^{1}$College of Science, Shanghai University,
Shanghai, 200436, P. R. China} \affiliation{$^{2}$Theoretical
High-Energy Physics and Cosmology group, Department of Physics,
Faculty of Science, Chulalongkorn University, Bangkok 10330,
Thailand} \affiliation{$^{3}$Center of Theoretical Nuclear
Physics, National Laboratory of Heavy Ion Accelerator, Lanzhou
730000, P. R. China} \affiliation{$^{4}$Institute of Modern
Physics, Lanzhou, 730000, P. R. China}

\begin{abstract}
We consider the Randall-Sundrum brane-world model with bulk-brane energy
transfer where the Einstein-Hilbert action is modified by curvature correction
terms: a four-dimensional scalar curvature from induced gravity on the brane,
and a five-dimensional Gauss-Bonnet curvature term. It is remarkable that
these curvature terms will not change the dynamics of the brane universe at
low energy. Parameterizing the energy transfer and taking the dark radiation
term into account, we find that the phantom divide of the equation of state of
effective dark energy could be crossed, without the need of any new dark
energy components. Fitting the two most reliable and robust SNIa datasets, the
182 Gold dataset and the Supernova Legacy Survey (SNLS), our model indeed has
a small tendency of phantom divide crossing for the Gold dataset, but not for
the SNLS dataset. Furthermore, combining the recent detection of the SDSS
baryon acoustic oscillations peak (BAO) with lower matter density parameter
prior, we find that the SNLS dataset also mildly favors phantom divide crossing.

\end{abstract}
\date{\today}
\maketitle

\section{Introduction}

Numerous cosmological observations have confirmed that the universe is
undergoing accelerated expansion. This phenomenon was not predicted by
conventional cosmology governed by general relativity with the known matter
constituents. To explain the cosmic acceleration, mysterious dark energy was
proposed. There are many dark energy models, which can be distinguished in the
value and variation of the equation of state (EoS) $w$ during the evolution of
the universe. The cosmological constant is the simplest candidate of dark
energy, whose EoS $w=-1$ is located at a central position among dark energy
models. For quintessence \cite{Caldwell}, Chaplygin gas \cite{Kamenshchik} and
holographic dark energy models \cite{LiMiao}, $w$ always stays bigger than -1.
The simplest model with $w$ $<-1$ is the phantom models \cite{Calwell} but
which will violate not only the Weak Energy Condition but also Null Energy
Condition. In general, dark energy can evolve with the change of the EoS
$w>-1$ to $w<-1$ (or vice versa). This transition is called \textquotedblleft
the phantom divide crossing\textquotedblright.

By current estimates, it seems likely that the phantom divide crossing occurs
at the recent epoch \cite{WangY}, though many of these estimations are model
dependent. In particular, with an advantageous parametrization (the called CPL
parametrization) of dynamical dark energy \cite{Linde}, it has been found
\cite{Nesseris05,Nesseris07} that most of the observational probes indeed
mildly favor dynamical dark energy crossing the phantom divide at $z\sim0.25$.
Moreover, in the era of structure formation, a highly negative $w$ makes
negligible the undesirable dark energy to the total energy density. Hence even
if the observation still admits $w=-1$, it is still useful to construct a more
general framework that can permit $w<-1$ \cite{Bronnikov}. However, it has
been proved that the phantom divide crossing of dark energy described by the
minimum coupling single scalar field with general Lagrangian is either
unstable with respect to the cosmological perturbations or realized on the
trajectories of the measure zero \cite{Vikman}. So most dark energy models
with the phantom divide crossing either consist of multiple scalar fields with
at least one non-canonical phantom component \cite{ZhangFX} or must recourse
to extending gravity theory \cite{Boisseau,Sahni}. The former is usually
plagued by catastrophic UV instabilities, however it can be regarded as an
effective field description following from an underlying theory with positive
energies \cite{Nojiri}. An interesting example where the UV pathologies are
absent in phantom models is given by \cite{Rubakov}. The latter, for example
the so called $1/R$ gravity \cite{Chiba}, is severely constrained by solar
system test and by cosmological observation even though it is interesting
theoretically. A novel model about $f(R)$ gravity which pass through all tests
is proposed by \cite{Starobinsky}. Other theories with the phantom divide
crossing are interacting holographic dark energy models \cite{WangB}, models
with interactions between dark matter and dark energy \cite{Chimento}, model
with a scale field coupled to the Gauss-Bonnet form \cite{Neupane}, model with
non-linear term of scalar curvature \cite{Srivastava}, and model considering
quantum effects \cite{Onemli}.

In another area of theoretical cosmology, the brane-world scenario has
received a lot of interest. The well known Randall-Sundrum (RS) brane-world
model \cite{Randall,Randall2}, inspired by D-brane ideology in string theory,
envisages that our four-dimensional universe is a 3 dimensional membrane
embedded in the five-dimensional bulk. The standard model particles are
confined in the brane while the gravity is free to propagate into the bulk.
This brane approach provides a new way of understanding the hierarchy between
the four-dimensional Planck scale and the electro-weak scale. Two important
generalizations of the RS model have been considered recently. The first is
inspired by superstring theory which suggests the Gauss-Bonnet (GB) curvature
corrections as the first and dominant quantum corrections to the
Einstein-Hilbert action for a ghost-free theory \cite{Zwiebach}. The combined
action in five dimensions gives the most general action with second-order
field equation, as shown by Lovelock \cite{Lovelock}. The DGP model suggests
the second curvature correction term to RS model, the four-dimensional scalar
curvature term. This induce gravity correction term can be interpreted as
arising from a quantum effect due to the interaction between the bulk
gravitons and the matter on the brane \cite{Dvali}. It is natural to study the
influence of the GB correction to the DGP model. Indeed, in certain
realization of string theory, the ghost-free GB term in the bulk action may
naturally lead to DGP induced gravity term on the brane boundary
\cite{Mavromatos}. It has been suggested that the combination of the two
curvature corrections may shed some light on the singularity problem in the
early universe \cite{Brown,Kofinas1}.

However, many brane-world models including the RS and GB types, produce only
ultra-violet modifications to general relativity, with extra-dimensional
gravity dominating at high energy. DGP model may manifest nontrivial low
energy extra-dimensional gravity, which results in a new branch DGP(+) with
late-time acceleration \cite{Deffayet}. Its generalization with the brane and
bulk cosmological constants \cite{Sahni} allows $w<-1$ and the branch DGP(-)
which has the right RS limit with $w>-1$, but none of these has the phantom
divide crossing behavior.

Another robust way that extra-dimension can affect the low energy evolution of
the universe is through the coupling between the brane and the bulk. This is
analogous to the coupled dark energy scenarios \cite{Ellis}, where the
late-time accelerating cosmological phase is characterized by a frozen ratio
of dark matter/dark energy as a result of the interaction of the dark matter
with other components, such as scalar fields. It has been proved that the
transfer of energy between the bulk and the brane may result in the RS model
with late-time acceleration \cite{Kiritsis} and even the phantom divide
crossing, provided that there is bulk matter \cite{Bogdanos} or an additional
dark energy on the brane \cite{Cai}. However, the former can not make
cosmological model which is independent of bulk dynamics, and the latter can
not have the phantom divide crossing by using only geometric effect. The
brane-bulk energy transfer has also been considered in DGP model for the
present universe as a global attractor using fix-points theory \cite{Kofinas}
and in many other different setups \cite{Hall,Tetradis}. The combined
curvature effect and bulk contents effect on the RS model have recently been
studied in \cite{Ahmad}.

The aim of the present work is to study the low-energy cosmological behavior
based on an extended RS(II) scenario \cite{Randall2} by considering brane-bulk
energy transfer and two additional curvature corrections. We obtain a closed
system of three equations which determines the parameters of the desired
Friedmann equation. In the low energy region, it is remarkable that the
curvature corrections terms do not change the dynamics of the system. By
parameterizing the energy transfer with the scale factor, we are able to
exactly solve the Friedmann equation. When including the effect of the dark
radiation, which is neglected in \cite{Cai,Bogdanos}, we show that the phantom
divide crossing may be achieved in the absence of additional dark energy
components on the brane. Furthermore, we fit the model to the two most
reliable and robust SNIa datasets, the new 182 Gold dataset \cite{Riess07} and
the first year Supernova Legacy Survey (SNLS) dataset \cite{Astier},
respectively complemented by the recent SDSS baryon acoustic oscillations peak
(BAO) dataset.

This letter is arranged as follows: In Sec. II, we establish the most general
brane world model with curvature correction terms and bulk-brane energy
transfer to describe the accelerated expansion. We investigate the equation of
state of effective dark energy and discuss the possibility of phantom divide
crossing. In Sec. III, we fit the model to the data from dark energy
observations. In the last section, we conclude with a brief summary.

\section{Brane world with curvature corrections and brane-bulk energy
transfer}

Let us consider a braneworld model. For convenience and without loss of
generality we choose the extra-dimension coordinates $y$ such that the brane
is located at $y=0$ and the bulk has $Z_{2}$ symmetry under the transformation
$y\rightarrow-y$. The most general action which incorporates the induced
gravity and Gauss-Bonnet corrections is \cite{Kofinas1}%
\begin{equation}
S=\frac{1}{2\kappa_{5}^{2}}\int d^{5}x\sqrt{-g}\{\mathcal{L}_{EH}%
+\alpha\mathcal{L}_{GB}\}+\frac{1}{2\kappa_{4}^{2}}\int_{brane}d^{4}%
x\sqrt{-\tilde{g}}\mathcal{L}_{IG}, \label{Sgrav}%
\end{equation}
where $g$ ($\tilde{g}$) and $\kappa_{5}$ ($\kappa_{4}$) are the bulk (brane)
metric and bulk (brane) gravitational constant, respectively. $L_{EH}%
=R-2\Lambda$ is the five-dimensional Einstein-Hilbert Lagrangian with negative
cosmological constant $\Lambda<0$. The Gauss-Bonnet curvature correction term
$L_{GB}$ is written as
\[
\mathcal{L}_{GB}=R^{2}-4R_{AB}R^{AB}+R_{ABCD}R^{ABCD}.
\]
We can define the Gauss-Bonnet coupling $\alpha$ through string energy scale
$g_{s}$ as $\alpha=\frac{1}{8g_{s}^{2}}$ . The induced-gravity Lagrangian
$L_{IG}=\tilde{R}-2\kappa_{4}^{2}\lambda$ consists of four-dimensional scale
curvature $\tilde{R}$ and brane tension $\lambda>0$. We can define the
induced-gravity crossover length scale by $r=\frac{\kappa_{5}^{2}}{\kappa
_{4}^{2}}$. For convenience, we will choose the unit that $\kappa_{5}=1$
throughout this letter. Note that by setting $r=\alpha=0$, we can recover the
RS model. The RS model with Gauss-Bonnet correction and the DGP model
correspond to the case with $r=0$ and $\alpha=0$, respectively.

By varying the action in Eq. (\ref{Sgrav}) with respect to the bulk metric, we
obtain the field equation%
\begin{equation}
G_{AB}+2\alpha H_{AB}=\left.  T_{AB}\right\vert _{total},
\label{field equation}%
\end{equation}
where $H_{AB}$ is the second order Lovelock tensor \cite{Lovelock}%
\[
H_{AB}=RR_{AB}-2R_{A}^{C}R_{BC}-2R^{CD}R_{ABCD}+R_{A}^{CDE}R_{BCDE}-\frac
{1}{4}g_{AB}\mathcal{L}_{GB}.
\]
The total energy-momentum tensor $\left.  T_{AB}\right\vert _{total}$ is
decomposed into bulk and brane components%
\[
\left.  T_{AB}\right\vert _{total}=\left.  T_{AB}\right\vert _{bulk}+\left.
T_{AB}\right\vert _{brane}\hat{\delta}(y).
\]
Here, we use the normalized Dirac delta function, $\hat{\delta}(y)=\sqrt
{\tilde{g}/g}\delta(y)$. The bulk component is
\[
\left.  T_{AB}\right\vert _{bulk}=-\Lambda g_{AB}+T_{AB},
\]
where $T_{AB}$ denotes any possible energy-momentum in the bulk. The brane
component is written as
\[
\left.  T_{AB}\right\vert _{brane}=-\lambda\tilde{g}_{AB}-r\tilde{G}%
_{AB}+\tilde{T}_{AB},
\]
where $\tilde{G}_{AB}$ arises from the scalar curvature in Eq. (\ref{Sgrav}).
The energy momentum tensor $\tilde{T}_{AB}$ represents matter on the brane
with energy density $\rho$ and constant equation of state parameter $w_{m}$.

The five-dimensional line element in the bulk is given by
\begin{equation}
ds^{2}=-n^{2}(t,y)dt^{2}+a^{2}(t,y)\gamma_{ij}dx^{i}dx^{j}+b^{2}(t,y)dy^{2}%
\end{equation}
where $\gamma_{ij}$ is a 3-dimensional maximally symmetric metric whose
spatial curvature is characterized by $k=0,\pm1$. In this letter, we are
interest in spatially flat brane $k=0$. We choose the coefficients $n(t,0)=1$
so that $t$ is the proper time along the brane. For simplicity, we assume that
the fifth dimension is static $\dot{b}=0$ and we set $b=1$.

To determine the Friedmann equation on the brane, we impose the junction
condition for a braneworld in Gauss-Bonnet gravity. For later use, we define
\[
\Phi=\frac{1}{n^{2}}\left(  \frac{\dot{a}}{a}\right)  ^{2}-\frac{a^{\prime2}%
}{a^{2}},
\]
where prime and dot denote the derivative with respect to $y$ and $t$,
respectively. The jump of the $(00)$ and $(ij)$ components of the field
equation Eq. (\ref{field equation}) across the brane gives%
\begin{equation}
2[1+\frac{8}{3}\alpha(H^{2}+\frac{\Phi_{0}}{2})]\frac{a_{+}^{\prime}}{a_{0}%
}=rH^{2}-u, \label{ap}%
\end{equation}%
\begin{equation}
2\frac{n_{+}^{\prime}}{n_{0}}+4\frac{a_{+}^{\prime}}{a_{0}}+8\alpha\frac
{n_{+}^{\prime}}{n_{0}}\Phi_{0}+16\alpha\frac{a_{+}^{\prime}}{a_{0}}%
(H^{2}+\dot{H})=3r(H^{2}+\frac{2}{3}\dot{H})+v, \label{np}%
\end{equation}
where $2a_{+}^{\prime}=-2a_{-}^{\prime}$ and $2n_{+}^{\prime}=-2n_{-}^{\prime
}$ are the discontinuities of the first derivatives. $H=\frac{\dot{a}_{0}%
}{a_{0}}$ is the Hubble constant on the brane and $\Phi_{0}=\Phi(t,0)$. We
define
\[
u=\frac{1}{3}(\rho+\lambda),\;v=\left(  w_{m}\rho-\lambda\right)  .
\]
One can show that Eq. (\ref{ap}) can be written in terms of the square of
Hubble parameter $H^{2}$%
\begin{equation}
4[1+\frac{8}{3}\alpha(H^{2}+\frac{\Phi_{0}}{2})]^{2}(H^{2}-\Phi_{0}%
)=[rH^{2}-u]^{2}. \label{junction}%
\end{equation}
Once we know the behaviors of $\Phi_{0}$ and $\rho$, Eq. (\ref{junction}) can
be solved and its solution will give the desired Friedmann equation.

Let us consider $\rho$ first. Since the left-hand side of Eq.
(\ref{field equation}) is divergence-free, the total energy momentum tensor is
conserved in the bulk $\nabla_{A}\left.  T_{B}^{A}\right\vert _{total}=0$. Its
zero component is%
\[
\dot{T}_{0}^{0}+3\frac{\dot{a}}{a}(T_{0}^{0}-T_{1}^{1})-(\frac{n^{\prime}}%
{n}+3\frac{a^{\prime}}{a})T_{0}^{5}+T_{0}^{5\prime}-[\dot{\rho}+3(1+w_{m}%
)H\rho]\delta(y)=0.
\]
Integrating around $y=0$ and using the $Z_{2}$ symmetry ($T_{0}^{5}%
(t,+0)=-T_{0}^{5}(t,-0)$), we can determine the evolution of $\rho$%
\begin{equation}
\dot{\rho}+3(1+w_{m})H\rho=2T_{05}. \label{ro}%
\end{equation}
This implies that the energy conservation law on the brane is broken. Another
method to obtain the semi-energy conservation law Eq. (\ref{ro}) is to solve
the junction conditions Eq. (\ref{ap}) and Eq. (\ref{np}). Their exact
solutions are complicated but we can make further simplification by assuming
$\alpha$ to be small. This is a reasonable assumption since we are interested
in the effect of GB correction on the late-time evolution of the universe. The
solutions of the Eq. (\ref{ap}) and Eq. (\ref{np}) up to the first order in
$\alpha$ are%
\begin{equation}
\frac{a_{+}^{\prime}}{a_{0}}=\frac{1}{2}\left(  rH^{2}-u\right)  +\frac{1}%
{6}\left(  rH^{2}-u\right)  \left[  r^{2}H^{4}+u^{2}-2H^{2}\left(
6+ru\right)  \right]  \alpha, \label{ap1}%
\end{equation}%
\begin{align}
\frac{n_{+}^{\prime}}{n_{0}}  &  =\frac{1}{2}\left(  2u+v+rH^{2}+2r\dot
{H}\right)  +\frac{1}{6}\{r^{3}H^{6}+3rH^{4}(-4+2ru+rv+2r^{2}\dot
{H})\label{ap2}\\
&  +u[u(8u+3v)+6(4+ru)\dot{H}]-3H^{2}[4v+u(8+5ru+2rv)+4r(4+ru)\dot{H}%
]\}\alpha.\nonumber
\end{align}
Note that Eq. (\ref{ap}) is a cubic equation for the discontinuity
$\frac{a_{+}^{\prime}}{a_{0}}$ which has one real solution, the other two are
complex. Since we require our cosmological equations to have the right
$\alpha\rightarrow0$ limit \cite{Germani}, we take only the real solution.
Substituting Eq. (\ref{ap1}) and Eq. (\ref{ap2}) into the $(05)$ component of
the field equation Eq. (\ref{field equation})%
\begin{equation}
3(\frac{n^{\prime}}{n}\frac{\dot{a}}{a}-\frac{\dot{a}^{\prime}}{a}%
)(1+4\alpha\Phi)=T_{05}, \label{05}%
\end{equation}
we recover (up to the first order of $\alpha$) the same semi-conservation law
Eq. (\ref{ro}).

Let us consider the function $\Phi$. For brane cosmology with Gauss-Bonnet
correction, $\Phi$ is just the first integral satisfying the constraint
equation%
\[
\Phi+2\alpha\Phi^{2}=\frac{\Lambda}{6}+\frac{C}{a^{4}},
\]
where $\frac{C}{a^{4}}$ is the dark radiation term \cite{Binetruy}. We point
out an important observation that the above first integral can be recovered
from a general differential equation with nontrivial $T_{05}$ and $T_{55}$%
\begin{equation}
\dot{\psi}+4\frac{\dot{a}}{a}\psi+4T_{05}\frac{a^{\prime}}{a}+4\frac{\dot{a}%
}{a}(T_{55}-\Lambda)=0, \label{chi}%
\end{equation}
where%
\[
\psi=6(\Phi+2\alpha\Phi^{2}).
\]
Notice that we only consider the solution having the right $\alpha
\rightarrow0$ limit%
\begin{equation}
\Phi=\frac{-3+\sqrt{3}\sqrt{3+4\alpha\psi}}{12\alpha}. \label{fp}%
\end{equation}
The differential equation Eq. (\ref{chi}) is obtained by substituting
$\frac{n^{\prime}}{n}$ from Eq. (\ref{05}) into the $(55)$ component of the
field equation Eq. (\ref{field equation})%
\[
3\left(  \frac{a^{\prime}}{a}\frac{n^{\prime}}{n}-\frac{1}{n^{2}}\frac
{\ddot{a}}{a}\right)  (1+4\alpha\Phi)-3\Phi=T_{55}-\Lambda.
\]
Note that for vanishing $T_{05}$ and $T_{55}$, the constraint equation Eq.
(\ref{chi}) gives a bulk equation. But for generic values of $T_{05}$, the
constraint equation depends on the discontinuity of the first derivative on
the brane. Moreover, we point out that the physical meaning of $\psi$ can be
understood as the effective bulk cosmological constant. To see this, we
rewrite Eq. (\ref{chi}) as%
\begin{equation}
\dot{\psi}_{1}+4\frac{\dot{a}}{a}\psi_{1}+4T_{05}\frac{a^{\prime}}{a}%
+4\frac{\dot{a}}{a}T_{55}=0 \label{psi99}%
\end{equation}
where $\psi_{1}=\psi-\Lambda$ takes the role as the correction to the bulk
cosmological constant. We can easily see that, for $T_{05}=T_{55}=0$, the only
correction to bulk cosmological constant $\Lambda$ is the dark radiation term
$\sim a^{-4}$. For a nontrivial bulk, the term $4T_{05}\frac{a^{\prime}}%
{a}+4HT_{55}$ in Eq. (\ref{chi}) will give further corrections to $\Lambda$.
We will come back to discuss this point later.

Generally, Eq. (\ref{junction}) has three solutions of $H^{2}$, but only two
of them are left for small $\alpha$:
\begin{equation}
H^{2}=\frac{6+3ru\mp\phi}{3r^{2}}+\frac{\alpha}{9r^{4}\phi}\{32[\mp
9r^{2}u+6ru\phi\mp2\phi(\mp6+\phi)]\pm8r^{2}(\pm18+3ru+\phi)\psi\pm r^{4}%
\psi^{2}\}, \label{H2}%
\end{equation}
where we have used Eq. (\ref{fp}) and $\phi=\sqrt{6}\sqrt{6+6ru-r^{2}\psi}$.
These two branches recover the two branches of DGP model when $\alpha
\rightarrow0$. One can find three equations (\ref{ro}), (\ref{chi}) and
(\ref{H2}) consisting of a closed system for variables $\rho$, $\Phi_{0}$ and
$H$, which determines the evolution of the universe (provided the bulk
energy-momentum tensor is known). However, these equations are entangled in a
complicated way. In the following, we will seek the simplest low-energy
effective theory which can be analytically solved.

We would like to make some assumptions here. First, we consider constraints on
$\rho$ and $H^{2}$. In low energy region, we can assume%
\begin{equation}
\rho\ll\lambda,\;\rho\ll\frac{\Lambda}{\lambda},\;\rho\ll\frac{\Lambda}%
{\alpha\lambda^{3}},\;H^{2}\ll\frac{\lambda}{r}. \label{low energy}%
\end{equation}
This low energy region is self-consistent and can be easily realized. The
simplest example is the case where the brane tension and the bulk cosmological
constant are very large. Moreover, it should be noticed that in the
$\alpha\rightarrow0$ and $r\rightarrow0$ limits Eq. (\ref{low energy}) is
consistent with RS low energy region%
\begin{equation}
\rho\ll\lambda,\;H\ll\lambda,\;\lambda\rho\sim H^{2},\;\Lambda\sim\lambda^{2}
\label{RS}%
\end{equation}
where the last constraint comes from the well known RS fine-tuning mechanics.
Noticing that the present Hubble rate $H_{0}$ is much smaller than the string
energy scale $g_{s}$ and $H^{2}/H_{0}^{2}$ is close to $1$ at late time, we
have $\alpha H^{2}\ll1$. Hence the region $\rho\ll\frac{\Lambda}{\alpha
\lambda^{3}}$ does not suggest new energy constraint beyond Eq. (\ref{RS}).
Similar consideration can be applied to the $rH^{2}\ll\lambda$ constraint in
Eq. (\ref{low energy}) if we set $rH\sim1$ in DGP(+) brane \cite{Brown}.
Second, to derive a cosmological system that is largely independent of the
bulk dynamics, usually one can assume that on the brane the contribution of
$T_{55}$ relative to the bulk vacuum energy is much less important than the
brane matter relative to the brane vacuum energy, or schematically
$\frac{T_{55}}{\Lambda}\ll\frac{\rho}{\lambda}$. Thus, $T_{55}$ can be omitted
in the low energy region. Third, the semi-conservation law Eq. (\ref{ro})
implies $T_{05}\precsim H\rho$ in general. Fourth, at late time, the dark
radiation may be assumed to be much smaller than bulk cosmological constant.
By taking in to account the above assumptions, we find that Eq. (\ref{psi99})
can be simplified as%
\begin{equation}
\dot{\psi}_{1}+4H\psi_{1}+4T_{05}A=0, \label{chi11}%
\end{equation}
where $A$ is a constant defined by%
\[
A=-\frac{\lambda}{6}(1+\frac{\lambda^{2}}{27}\alpha).
\]
It should be noticed that $T_{05}A\ll H\Lambda$. Returning to the discussion
about the correction to the bulk cosmological constant. We can now conclude
that the correction to the bulk cosmological constant $\psi_{1}$ is much
smaller than the bulk cosmological constant, $\psi_{1}\ll\Lambda$. This is
reminiscent of $\rho\ll\lambda$ on the brane, and can be understood as
bulk-brane duality. Thus, Eq. (\ref{H2}) can be expanded in terms of $\rho$
and $\psi_{1}$ (keeping only the first order terms):%
\begin{equation}
H^{2}=B\rho+C\psi_{1}+D, \label{H21}%
\end{equation}
where we absorb $\lambda$ and $\Lambda$ into the constant coefficients%
\begin{align}
B  &  =\frac{E\mp6}{3rE}+\frac{2\alpha}{9r^{3}E^{3}}\{32(E\mp6)E^{2}%
-9r[\pm8r\Lambda(8+3r\lambda)\pm32\lambda(4+r\lambda)\pm3r^{2}\Lambda
^{2}]\},\nonumber\\
C  &  =\pm\frac{1}{E}+\frac{\alpha}{E^{3}r^{2}}[-96(E\lambda\mp6)-32r\lambda
(E\mp9)+8r^{2}\Lambda(2E\mp27)\mp3r^{4}\Lambda^{2}],\nonumber\\
D  &  =\frac{6\mp E+r\lambda}{3r^{2}}+\frac{\alpha}{9Er^{4}}\{64Er\lambda
\mp64E(6\mp E)\mp32r^{2}\lambda^{2}\pm8r^{2}[\mp\Lambda(-18\mp E+r\lambda)\pm
r^{4}\Lambda^{2}]\}, \label{BCD}%
\end{align}
with $E=\sqrt{6(6+2r\lambda-r^{2}\Lambda)}$.

Observing the three equations Eq. (\ref{ro}), Eq. (\ref{chi11}), and Eq.
(\ref{H21}), we find that the curvature corrections only affect the constants
$A,B,C,D,E$, and they will not vanish in the $\alpha\rightarrow0$ and
$r\rightarrow0$ limits. This remarkable result immediately tells us that the
curvature corrections will not change the form of the desired Friedmann
equation, and the corresponding evolution of the universe.

To obtain the explicit solution we need to know the form of energy transfer
$T_{05}$. Unfortunately, it is not yet available and obviously depends on the
mechanism which produces the energy transfer. We consider the ansatz
$T_{05}=THa^{v}$ ($T$ is a constant) which was used in \cite{Cai,Bogdanos} for
RS brane world. A justifcation for the ansatz has been analyzed in
\cite{Bogdanos} through a simple model where the bulk content is a
relativistic fluid, slowly moving along the fifth dimension. Recently, the
ansatz has also been used in \cite{Bogdanos1}. Then we find
\begin{equation}
\rho=Fa^{-3}+\frac{2T}{3+v}a^{v} \label{ro2}%
\end{equation}
where we take $w_{m}=0$ for dark matter, and%
\begin{equation}
\psi_{1}=Ga^{-4}+\frac{2T\lambda(27+\alpha\lambda^{2})}{81(4+v)}a^{v}
\label{chi12}%
\end{equation}
where $G$ is an integration constant. Notice that the last term in Eq.
(\ref{ro2}) denotes the energy flow into ($\frac{2T}{3+v}>0$) or out of
($\frac{2T}{3+v}<0$) the brane. Substituting Eq. (\ref{ro2}) and Eq.
(\ref{chi12}) into Eq. (\ref{H21}), we have%
\[
H^{2}=\Omega_{0m}a^{-3}+\Omega_{0v}a^{v}+\Omega_{0d}a^{-4}+D,
\]
where%
\[
\Omega_{0m}=BF,\;\Omega_{0v}=B\frac{2T}{3+v}+C\frac{2T\lambda(27+\alpha
\lambda^{2})}{81(4+v)},\;\Omega_{0d}=CG.
\]
To impose the vanishing effective cosmological constant on the brane, we
choose $D=0$. It just recovers RS fine-tuning when $\alpha\rightarrow0$.
Finally, we can write%
\begin{equation}
H^{2}=\Omega_{0m}a^{-3}+\Omega_{0d}a^{-4}+\Omega_{0v}a^{v}. \label{fhh}%
\end{equation}
This is similar to the well known result in RS model where the Friedmann
equation can be generalized by adding the term depending on the brane-bulk
energy flow. The curvature corrections only affect the coefficients
$\Omega_{0m}$, $\Omega_{0d}$ and $\Omega_{0v}$. However, it should be pointed
out that, they may play an important role, for example, when the constant $G$
in dark radiation term and the parameter $T$ which characterizes the
brane-bulk energy exchange are not big enough to provide desired nontrivial
cosmological behavior. In particular, the dark radiation term is usually
neglected if one considers the large scale factor at late time. In RS model,
the reason that we can keep it is that indeed we do not know the constant $G$
which reflects the bulk geometry (Notice that it takes the role of the bulk
black hole mass in a Schwarzschild-AdS$_{5}$ geometry \cite{Germani}). In our
model, furthermore, the curvature corrections embodied in constant $C$ may
strengthen the need for keeping the dark radiation term. For an explicit
example, we assume $G\ll6F\lambda a$ so that the dark radiation term can be
omitted $\Omega_{0d}a^{-4}\ll\Omega_{0m}a^{-3}$ in RS case where $C=\frac
{1}{6}$, $B=\lambda$. For the dark radiation term to be important $\Omega
_{0d}a^{-4}\sim\Omega_{0m}a^{-3}$, we need $G\sim\frac{BFa}{C}\ll6F\lambda a$
i.e. $\frac{B}{C}\ll6\lambda$. Considering the induced gravity correction and
using lower branch for simplicity, we find that the condition is satisfied
when $r\gg\frac{38\lambda}{54\lambda^{2}+\Lambda}$, which does not violate the
low energy region Eq. (\ref{low energy}). Another example can be given by
considering the relationship between the accelerated expansion and the
brane-bulk energy flow. The accelerated expansion is characterized by the
deceleration parameter $q=-\frac{\ddot{a}/a}{H^{2}}$. Using $a=\frac{a_{0}%
}{1+z}$ and taking $\left.  H^{2}\right\vert _{z=0}=1$, we can rewrite the
Friedmann equation as%
\[
H^{2}=\Omega_{0m}(1+z)^{3}+\Omega_{0v}(1+z)^{-v}+(1-\Omega_{0m}-\Omega
_{0v})(1+z)^{4}%
\]
where we have absorbed $a_{0}$ into $\Omega_{0m}$ and $\Omega_{0v}$. The
deceleration parameter can be given as%
\[
q=-1+\frac{1}{2}\frac{d\log H^{2}}{d\log(1+z)}.
\]
Now we will omit the terms which decrease more quickly than matter density at
late time, then the dark radiation term is absent and one can find that the
present accelerated expansion $\left.  q\right\vert _{z=0}<0$ needs $v>-3$ and
$\Omega_{0v}>\frac{1}{3+v}>0$. This suggests that in RS case (noticing
$\lambda>0$ in RS(II) brane world), the bulk energy must flow into the brane
(the later term in Eq. (\ref{ro2}) is positive, i.e. $T>0$). However, if we
consider the curvature correction (still the induced gravity correction and
using the lower branch), we find that accelerated expansion may be achieved
when the bulk energy flows out of the brane, which only needs $\left\vert
\frac{B}{C}\right\vert <\frac{v+3}{3(v+4)}\lambda$ that can be easily realized.

\begin{figure}[ptb]
\begin{center}
\includegraphics[
height=2in, width=3in ]{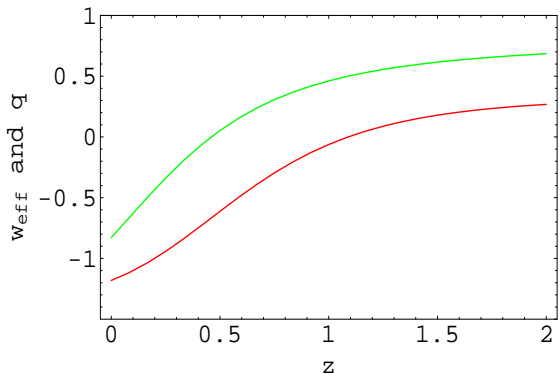}
\end{center}
\caption{The EoS $w_{eff}$ (red line) with prior $z_{T}=0.2$, $v=1$ and
deceleration factor $q$ (green line) with prior $z_{T}=0.2$, $v=1$,
$\Omega_{0m}=0.25$ versus redshift $z$. One can find that $w_{eff}$ crosses
-1, and the $q$ crosses 0 at $z\sim0.5$ and the order of magnitude $\left.
q\right\vert _{z=0}\sim-1$.}%
\label{fig1}%
\end{figure}

Let us consider whether or not this model is permitted to cross the phantom
divide and what is favored by fitting the model to observational datasets. We
find that the combination of the last two terms in Eq. (\ref{fhh}) may achieve
the phantom divide crossing, without the need of any other dark energy
components. The EoS of effective dark energy can be written as \cite{Linder}%
\[
w_{eff}=-1+\frac{1}{3}\frac{d\log[H^{2}-\Omega_{0m}(1+z)^{3}]}{d\log(1+z)}.
\]
It is easy to see that $w_{eff}$ is a constant if the dark radiation term is
omitted $\Omega_{0v}=1-\Omega_{0m}$ or the brane-bulk energy transfer is
trivial $\Omega_{0v}=0$. But if both of them are taken into account, we may
achieve the phantom divide crossing in very broad parameter space. Explicitly,
let us assume $w_{eff}=-1$ at $z_{T}$, then%
\[
\left.  w_{eff}\right\vert _{z=0}=-1+\frac{4v[1-(1+z_{T})^{4+v}]}%
{3[v+4(1+z_{T})^{4+v}]}.
\]
When $z_{T}>0$, one can find $\left.  w_{eff}\right\vert _{z=0}<-1$ if $v>0$.
This implies that the brane-bulk energy flow increases with the expansion of
the universe. For an explicit example of the phantom divide crossing and
current accelerated expansion, see Fig. (\ref{fig1}).

\section{Observation of the phantom divide crossing}

Now we use the observational datasets to test our cosmological model. We will
use the new 182 gold supernova Ia data ($0<z<1.76$) and the first year
Supernova Legacy Survey (SNLS) dataset ($0<z<1$), combined with the recent BAO
measurement from SDSS to fit our model. Recently these data have been widely
used in the fittings of different braneworld cosmological models \cite{aa}.
There are other dark energy observational probes, including the 3-year WMAP
CMB shift parameter \cite{Spergel}, the X-ray gas mass fraction in clusters
\cite{Allen} and the linear growth rate of perturbations at $z=0.15$ as
obtained from the 2dF galaxy redshift survey \cite{Hawkins}. However, since
our model is effective at low energy, we will not use the 3-year WMAP CMB
shift parameter focusing on the high redshift region. Besides, for simplicity,
we do not adopt other probes of dark energy which have large relative errors
compared with SnIa, SNLS, CMB and BAO probes \cite{Nesseris07}.

For the supernova Ia data, the measured quantity is the bolometric magnitude
$m$%
\[
m=\bar{M}+5\log_{10}(D_{L})
\]
where $\bar{M}$ is the Hubble-parameter-free absolute magnitude%
\[
\bar{M}=M-5\log_{10}(\frac{H_{0}^{-1}}{Mpc})+25.
\]
$M$ is the absolute magnitude, and%
\[
D_{L}=(1+z)\int_{0}^{z}dz^{\prime}\frac{H_{0}}{H(\Omega_{0m},\Omega_{0v},v)}%
\]
is the Hubble free luminosity distance ($H_{0}d_{L}/c$). The data points of
the Gold dataset are given after implementing correction for galactic
extinction, $K$-correction and light curve width-luminosity correction, in
terms of the distance modulus%
\[
u_{obs}^{Gold}(z_{i})\equiv m_{obs}^{Gold}(z_{i})-M.
\]
For SNLS datasets, also presents for each point, the stretch factor $s$ used
to calibrate the absolute magnitude and the rest frame color parameter $c$
which mainly measures host galaxy extinction by dust. Thus, the distance
modulus in this case depends apart from the absolute magnitude $M$, on two
additional parameters $\alpha$ and $\beta$
\[
u_{obs}^{SNLS}(z_{i})\equiv m_{obs}^{SNLS}(z_{i})-M+\alpha(s_{i}-1)-\beta
c_{i}\text{.}%
\]
Let us define the theoretical distance modulus%
\[
u_{th}(z_{i})\equiv m_{th}(z_{i})-M=5\log_{10}(D_{L})-5\log_{10}(\frac
{H_{0}^{-1}}{Mpc})+25\text{.}%
\]
We shall assume that the supernova Ia measurements come with uncorrelated
Gaussian errors $\sigma_{i}^{2}$ (including flux uncertainties, intrinsic
dispersion of SnIa absolute magnitude and peculiar velocity dispersion) in
which case the likelihood function is given by the $\chi^{2}$ distribution%
\begin{equation}
\chi^{2}(\Omega_{0m},\Omega_{0v},v)=\sum_{i}^{N}\frac{\left[  u_{obs}%
(z_{i})-u_{th}(z_{i})\right]  ^{2}}{\sigma_{i}^{2}} \label{chi2}%
\end{equation}
where $N=182$ for Gold datasets, and $N=115$ for SNLS.

For BAO measurement, we shall use the model independent measurement of the
parameter \cite{Alcaniz}%
\[
A=\Omega_{0m}^{\frac{1}{2}}\left(  \frac{H_{0}}{H}\right)  ^{\frac{1}{3}%
}\left[  \frac{1}{0.35}\int_{0}^{0.35}\frac{H_{0}}{H}dz\right]  =0.469\pm
0.017
\]
to construct an additional term in the $\chi^{2}$ equation Eq. (\ref{chi2})%
\[
\chi_{BAO}^{2}=\frac{\left[  A(\Omega_{0m},\Omega_{0v},v)-0.469\right]  ^{2}%
}{0.017^{2}}.
\]
The theoretical model parameters are determined by minimizing the $\chi^{2}$
and $\chi^{2}+\chi_{BAO}^{2}$.

\begin{figure}[ptb]
\begin{center}
\includegraphics[
height=2.6in, width=6in ]{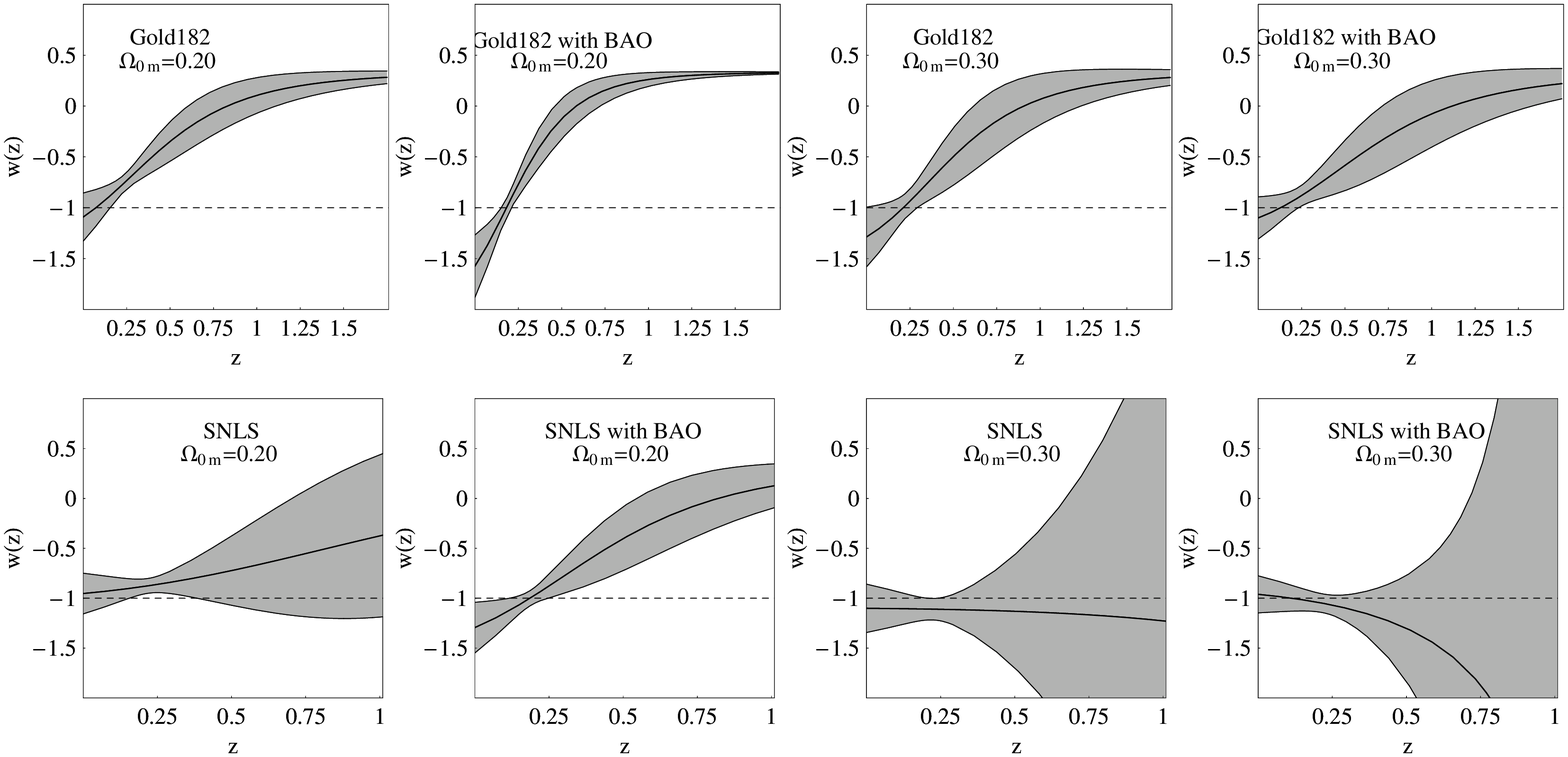}
\end{center}
\caption{The best fit form of $w_{eff}(z)$ for different dataset category for
both $\Omega_{0m}=0.2$ and $\Omega_{0m}=0.3$. The categories are: Gold182
dataset and it with BAO (row 1), SNLS and it with BAO (row 2). The dashed line
in each panels represents the phantom divide.}%
\label{fig2}%
\end{figure}

\begin{figure}[ptb]
\begin{center}
\includegraphics[
height=2.6in, width=6in ]{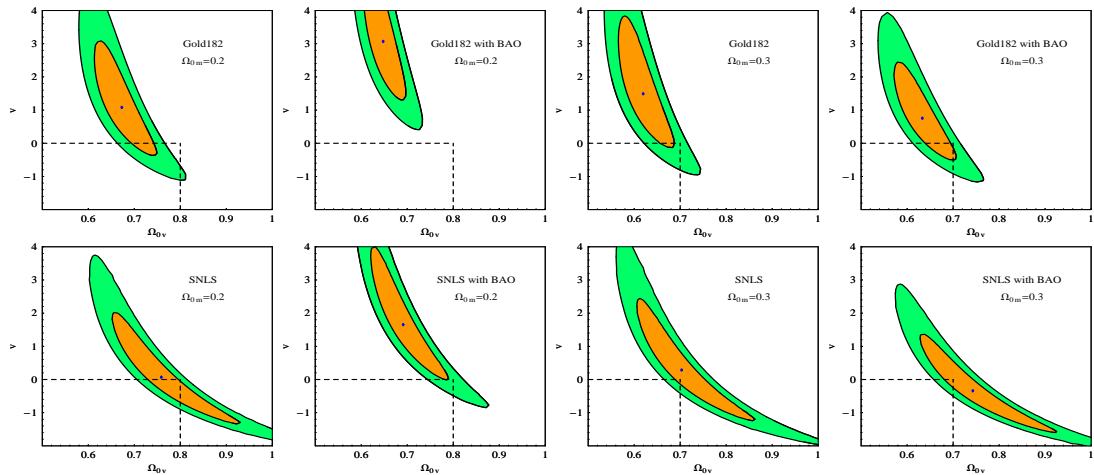}
\end{center}
\caption{The $68\%$ and $95\%$ confidence contours in the $\Omega_{0v}-v$
parameter space for each dataset category for both $\Omega_{0m}=0.2$ and
$\Omega_{0m}=0.3$. The cross point of dashed lines in each panels represents
$\Lambda$CDM.}%
\label{fig3}%
\end{figure}

\bigskip In order to investigate the dependence of the resulting best fits on
the prior of $\Omega_{0m}$, we will consider two cases $\Omega_{0m}$ = 0.2 and
$\Omega_{0m}$ = 0.3 instead of marginalizing over $\Omega_{0m}$. The range
between the two cases includes the current best fit value of $\Omega_{0m}$
based on WMAP and SDSS which is $\Omega_{0m}$ = 0.24$\pm$0.02 \cite{Tegmark}.
Considering the errors using the covariance matrix method \cite{Press}, we
show the best fit form of $w_{eff}$ for each dataset category in Fig.
(\ref{fig2}). The corresponding $\chi^{2}$ contours in the remaining two
parameter space is shown in Fig. (\ref{fig3}). 

\section{Summary}

We have studied the cosmological dynamics of the general RS brane world
scenario with brane-bulk energy transfer and two curvature correction terms.
For small Gauss-Bonnet coupling, we obtain a closed system of three equations
having two branches that correspond to the two branches in the DGP model. They
describe the evolution of the Hubble rate, the energy density and the
time-dependent effective bulk cosmological "constant". We find that in the low
energy region Eq. (\ref{low energy}), these two branches have equivalent
dynamics as that of the RS model with the brane-bulk energy exchange.
Furthermore, we have shown that the phantom divide crossing can be achieved
when the dark radiation term is presented and the brane-bulk energy flow
increases with the expansion of the universe, without the need of any
additional dark energy components on the brane \cite{Cai}, or bulk matter
\cite{Bogdanos}, or even exotic phantom material \cite{Calwell}. From the
fitting, it has been revealed that the model indeed has a small tendency of
phantom divide crossing for the Gold dataset, but not for the SNLS dataset.
This is consistent with the analysis in \cite{Nesseris05,Nesseris07} for CPL
parametrization of dynamical dark energy. We have further shown that the BAO
constraint with the lower matter density prior mildly changes the tendency of
SNLS dataset and favors the $w$ crossing $-1$.

\section*{Acknowledgment}

This work was supported by the National Natural Science Foundation of China
under Grant Nos. 10575068 and 10604024, the CAS Knowledge Innovation Project
Nos. KJcx.syw.N2, the Shanghai Education Development Foundation, and the
Natural Science Foundation of Shanghai Municipal Science Technology Commission
under grant Nos. 04ZR14059 and 04dz05905.

\end{document}